\documentclass{elsart}
\usepackage{epsfig}
\usepackage{graphicx}
\usepackage{amsmath}
\usepackage{amssymb}
\usepackage{amsfonts}

\begin{document}

\begin{frontmatter}
\title{Measurement of double beta decay of \nuc{100}{Mo}
to excited states in the NEMO~3 experiment}
\author[IReS]{R.~Arnold\corauthref{RA}} \ead{Roger.Arnold@IReS.in2p3.fr},
\author[LAL]{C.~Augier},
\author[INEL]{J.~Baker},
\author[ITEP]{A.S.~Barabash},
\author[LAL]{M.~Bongrand},
\author[CENBG]{G.~Broudin},
\author[JINR]{V.~Brudanin},
\author[INEL]{A.J.~Caffrey},
\author[JINR]{V.~Egorov}, 
\author[LAL]{A.I.~Etienvre},
\author[UM]{N.~Fatemi-Ghomi},
\author[CENBG]{F.~Hubert},
\author[CENBG]{Ph.~Hubert}, 
\author[CTU]{J.~Jerie},
\author[IReS]{C.~Jollet}, 
\author[LAL]{S.~Jullian},
\author[UCL]{S.~King},
\author[JINR]{O.~Kochetov},
\author[ITEP]{S.I.~Konovalov},
\author[JINR]{V.~Kovalenko},
\author[LAL]{D.~Lalanne},
\author[Maroc]{T.~Lamhamdi},
\author[CENBG]{F.~Leccia},
\author[LPC]{Y.~Lemi\`{e}re},
\author[LPC]{C.~Longuemare},
\author[CENBG]{G.~Lutter},
\author[CENBG]{Ch.~Marquet},
\author[LPC]{F.~Mauger},
\author[CENBG]{A.~Nachab},
\author[Saga]{H.~Ohsumi},
\author[CENBG]{F.~Perrot},
\author[CENBG]{F.~Piquemal},
\author[LSCE]{J.L.~Reyss},
\author[CENBG]{J.S.~Ricol},
\author[UCL]{R.~Saakyan},
\author[LAL]{X.~Sarazin},
\author[LAL]{L.~Simard},
\author[CU]{F.~\v{S}imkovic},
\author[JINR]{Yu.~Shitov},
\author[JINR]{A.~Smolnikov},
\author[UM]{S.~S\"{o}ldner-Rembold},
\author[CTU]{I.~\v{S}tekl},
\author[Jyva]{J.~Suhonen},
\author[MHC]{C.S.~Sutton},
\author[LAL]{G.~Szklarz},
\author[UCL]{J.~Thomas},
\author[JINR]{V.~Timkin},
\author[JINR]{V.~Tretyak},
\author[ITEP]{V.~Umatov},
\author[CTU]{L.~V\'{a}la},
\author[ITEP]{I.~Vanyushin},
\author[UCL]{V.~Vasiliev},
\author[Charles]{V.~Vorobel},
\author[JINR]{Ts.~Vylov}
\collab{NEMO Collaboration}

\address [IReS] {IPHC, IN2P3-CNRS et Universit\'e Louis Pasteur, 
  F-67037 Strasbourg, France}
\address [LAL] {LAL, IN2P3-CNRS et Universit\'e Paris-Sud, 
  F-91405 Orsay, France}
\address [INEL] {INEEL, Idaho Falls, ID83415, USA}
\address [ITEP] {Institute of Theoretical and Experimental Physics, 
  117259 Moscow, Russia}
\address [CENBG] {CENBG, IN2P3-CNRS et Universit\'e Bordeaux~I, 
  F-33170 Gradignan, France}
\address [JINR] {Joint Institute for Nuclear Research, 
  141980 Dubna, Russia}
\address [UM] {University of Manchester, M13 9PL Manchester,
  United Kingdom}
\address [CTU] {IEAP, Czech Technical University in Prague, 
  CZ-12800 Prague, Czech Republic}
\address [UCL] {University College London, 
  WC1E 6BT London, United Kingdom}
\address [Maroc] {USMBA, Fes, Maroc}
\address [LPC] {LPC, IN2P3-CNRS et Universit\'e de Caen, 
  F-14032 Caen, France}
\address [Saga] {Saga University, Saga 840-8502, Japan}
\address [LSCE] {LSCE, CNRS, F-91190 Gif-sur-Yvette, France}
\address [CU] {FMFI, Commenius University, SK-842 48 Bratislava, Slovakia}
\address [Jyva] {Jyv\"{a}skyl\"{a} University, 
  40351 Jyv\"{a}skyl\"{a}, Finland}
\address [MHC] {MHC, South Hadley, Massachusetts, MA01075, USA}
\address [Charles] {Charles University, Prague, Czech Republic}

\corauth [RA] {Corresponding author.}
\begin{abstract}
The double beta decay of $^{100}$Mo to the 0$^+_1$ and 2$^+_1$ excited
states  of $^{100}$Ru  is studied  using the  NEMO~3 data.   After the
analysis of 8024~h  of data the half-life for  the two-neutrino double
beta decay of  \nuc{100}{Mo} to the excited $0^+_1$  state is measured
to  be  T$^{(2\nu)}_{1/2}$ =
[5.7$^{+1.3}_{-0.9}$(stat)$\pm$0.8(syst)]$\cdot10^{20}$~y.
The signal-to-background ratio is equal to 3.
Information   about  energy  and  angular 
distributions of  emitted electrons is also obtained.  No evidence for
neutrinoless double beta  decay to the excited $0^+_1$  state has been
found.       The       corresponding      half-life      limit      is
T$^{(0\nu)}_{1/2}(0^+\rightarrow0^+_1)   >  8.9\cdot10^{22}$~y  (at
90\% C.L.).

The  search for the  double beta  decay to  the 2$^+_1$  excited state
has allowed the determination of limits on  the half-life for  the two
neutrino       mode      T$^{(2\nu)}_{1/2}(0^+\rightarrow2^+_1)      >
1.1\cdot10^{21}$~y  (at 90\%  C.L.)  and  for  the neutrinoless  mode
$T^{(0\nu)}_{1/2}(0^+\rightarrow2^+_1)  >  1.6\cdot10^{23}$~y  (at
90\% C.L.).
\end{abstract}

\end{frontmatter}


\section{Introduction}  
The  main
interest in double beta decay  is connected with the neutrinoless mode
$(0\nu\beta\beta)$ as a probe for physics beyond the Standard Model of
electroweak interactions.  Its existence is connected with fundamental
aspects of particle physics, for example, non-conservation of the 
lepton number, the  existence   and  nature  of  neutrino  mass,   
the  existence  of
right-handed currents in the electroweak interaction, the existence of
a massless Goldstone boson (the Majoron) and the existence
of supersymmetry~\cite{FAE01,Elliot2002,VER02}).

Interest  in neutrinoless  double beta  decay has  seen  a significant
rebirth in  recent years after the evidence  for neutrino oscillations
was obtained  from the results  of atmospheric~\cite{SOB01}  and
solar~\cite{CLE98,FUK01,GAV01,ALT00,AHM02}  neutrino  experiments
(see,  for 
example, the discussions in~\cite{HOL02,BAH02,MAL03}).
This observation of oscillations was recently confirmed by the KamLAND
experiment with reactor antineutrinos~\cite{EGU03} and by the new SNO
result~\cite{AHM04}.   These results  are  impressive proof  that
neutrinos  have a  non-zero mass.   However, the  experiments studying
neutrino oscillations are not sensitive  to the nature of the neutrino
mass (Dirac or Majorana?)  and  provide no information on the absolute
scale  of the neutrino  masses, since  such experiments  are sensitive
only to  neutrino mass-squared difference $\Delta m^{2}$.
The  detection and  study of $0\nu\beta\beta$ decay is the only practical
method to clarify the following characteristics of the neutrino
(see   discussions  in~\cite{PAS03,PAS04,BIL04,PAS05}):  (i)
neutrino nature (is the neutrino a Dirac or a Majorana particle?), (ii)
absolute  neutrino mass  scale (a  measurement or  a limit  on m$_1$),
(iii)  the  type of  neutrino  mass  hierarchy  (normal, inverted,  or
quasi-degenerate), (iv) CP violation in the lepton sector 
(measurement of the Majorana CP-phases, which requires
high precission neutrino oscillation data and reliable
nuclear matrix elements).

In connection with the $0\nu\beta\beta$ decay, the detection of double
beta decay with emission of two neutrinos $(2\nu\beta\beta)$, which is
an allowed process of second order in the Standard Model, provides the
possibility for experimental determination  of the 
nuclear matrix  elements (NME) involved
in the double beta decay  processes.  This leads to the development of
theoretical  schemes for nuclear  matrix-element calculations  both in
connection   with  the   $2\nu\beta\beta$  decays   as  well   as  the
$0\nu\beta\beta$ decays (see, for example, \cite{Rodin}).  At present,
$2\nu\beta\beta$  to  the ground  state  of the  final
daughter   nucleus  has been   measured  for   ten   nuclei:  \nuc{48}{Ca},
\nuc{76}{Ge},      \nuc{82}{Se},      \nuc{96}{Zr},     \nuc{100}{Mo},
\nuc{116}{Cd},   \nuc{128}{Te},   \nuc{130}{Te},   \nuc{150}{Nd}   and
\nuc{238}{U}  (a   review  of  results  is  given,   for  example,  in
Refs.~\cite{BAR02,BAR04}).

The $\beta\beta$  decay can proceed through transitions  to the ground
state as  well as to various  excited states of  the daughter nucleus.
Studies of  the latter transitions  allow supplementary
information about  $\beta\beta$ decay to be obtained.  
Because  of smaller transition
energies,  the  probabilities  for $\beta\beta$-decay  transitions  to
excited  states  are   substantially  suppressed  in  comparison  with
transitions to the ground state.  The first exclusive studies of
$\beta\beta-\gamma$ coincidence experiments have been done on $^{100}$Mo
about 15 year ago~\cite{Kudomi1992}.
But as it was shown in Ref.~\cite{BAR90}, by
using   low-background  facilities   utilising  High Purity Germanium
(HPGe)   detectors,  the 
$2\nu\beta\beta$ decay  to the $0^+_1$  level in the  daughter nucleus
may  be detected for  such nuclei  as \nuc{100}{Mo},  \nuc{96}{Zr} and
\nuc{150}{Nd}.  In this case  the energies involved in the
$\beta\beta$   transitions  are   large  enough   (1903,  2202   and
2627~keV, 
respectively),  and  the  expected  half-lives  are of  the  order  of
$10^{20} - 10^{21}$~y.  The  required sensitivity was only reached for
\nuc{100}{Mo}  (the  transition  was  detected  in  three  independent
experiments~\cite{BAR95,BAR99,BRA01}) and \nuc{150}{Nd}~\cite{BAR04a}.
Recently    additional    isotopes,    \nuc{82}{Se},    \nuc{130}{Te},
\nuc{116}{Cd}  and  \nuc{76}{Ge},  have  also become  of  interest  to
studies of the $2\nu\beta\beta$ decay to the $0^+_1$ level (see review
in Refs.~\cite{BAR04,BAR00}).

\begin{figure}[tb]
  \begin{center} 
    \includegraphics[height=7cm]{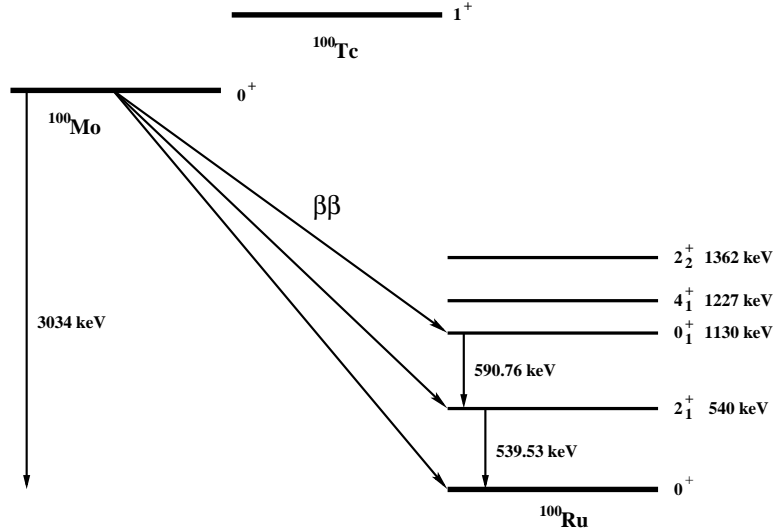} 
  \end{center} 
  \caption{A double beta decay scheme of \nuc{100}{Mo}
    showing the decay to the ground and excited states of
    \nuc{100}{Ru}.
   The latter transitions are followed by de-excitation $\gamma$-rays.
}
  \label{fig:mo100_decay_scheme} 
\end{figure}
Theoretical estimates of the $2\nu\beta\beta$ decay to a $2^+$ excited
state have  shown that for  a few nuclei  (\nuc{82}{Se}, \nuc{96}{Zr},
\nuc{100}{Mo}, and \nuc{130}{Te}) the  half-lives can be $\sim~10^{22}
- 10^{23}$~y~\cite{SUH98,CIV99}.  This would mean that the detection of such
decays becomes possible using the present and new installations in the
future.

It is very important to note  that in the framework of QRPA models the
behaviour  of   NME(2$\nu$) parameter $g_{pp}$ is  completely
different for transitions to the  ground and excited ($2^+$ and $0^+$)
states~\cite{SUH98,AUN96}.   This is why  the decay to  excited states
may probe different aspects of the calculation method than the decay
to the  ground states.  The $0\nu\beta\beta$ transition  to excited
states of  daughter nuclei, which provides a clear-cut signature  of 
such  decays, is
worthy of special  note here: in addition to two electrons with fixed
total energy, there appears one  ($0^+ - 2^+$ transition) or two ($0^+
-  0^+$ transition) photons,  their energies  being strictly  fixed as
well.  In  a hypothetical experiment  detecting all decay products  with a
high efficiency and a high  energy resolution, the background could be
reduced nearly  to zero.  Possibly  this idea will be  used in
future experiments featuring a large mass of the substance under study
(as  mentioned    in   Refs.~\cite{BAR04,BAR00,SUH00}).     In
Ref.~\cite{SIM02} it  was mentioned that detection  of this transition
will   give    us  the additional   possibility    to   distinguish   the
$0\nu\beta\beta$   mechanisms (the light and heavy Majorana neutrino
exchange mechanisms, the trilinear R-parity breaking mechanisms etc.).
So   the   search  for   $\beta\beta$  
transitions to the excited states has its own special interest.

The     $^{100}$Mo     decay     scheme     is    shown     in     
Fig.~\ref{fig:mo100_decay_scheme}.  The $\beta\beta$ decay  to the
0$^+_1$ excited 
state of  $^{100}$Ru is accompanied  by the simultaneous emission of
two electrons 
and two  $\gamma$-rays of the energies of 540~keV and 590~keV.  The emission of
two electrons and of one $\gamma$-ray of  energy 540~keV is the signature of the
double beta decay to the 2$^+_1$ excited state.


\section{NEMO~3 experiment}

\begin{figure}
  \begin{center} 
    \includegraphics[height=10cm]{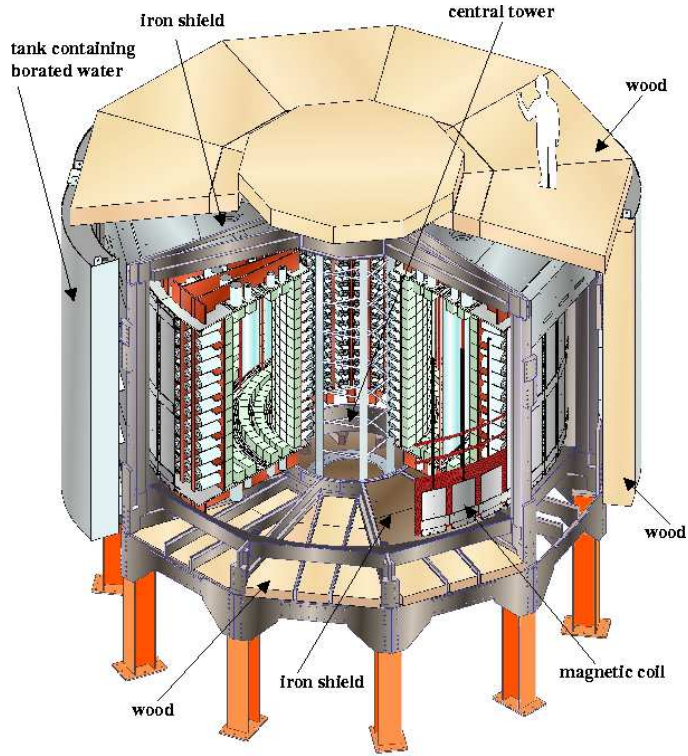} 
  \end{center}
  \caption{A schematic view of the NEMO~3 detector.} 
  \label{fig:nemo3-detector}  
\end{figure} 
NEMO~3 is a currently running experiment searching for
neutrinoless   double   beta   decay.
The philosophy behind the NEMO~3detector is the direct detection of the
two electrons from $\beta\beta$ decay by a tracking device and a
calorimeter~\cite{ARN05}. A schematic view of the NEMO~3 detector is
shown in Fig.~\ref{fig:nemo3-detector}. 
  The NEMO~3 detector  has three main components, 
a foil consisting
of different sources of double beta decay isotopes and copper, 
a tracker
made  of Geiger  wire cells  and  a calorimeter  made of  scintillator
blocks  with  PMT  readout,  surrounded  by a  solenoidal  coil.   The
detector has  the ability to discriminate between  events of different
types by  positive identification of charged tracks  and photons.  The
estimated  sensitivity of  this  detector for  $0\nu\beta\beta$ decay,  a
function of both  the detector design and the isotope  type, is of the
order of $10^{25}$~y, which  corresponds to an effective neutrino mass
$\langle m_\nu \rangle$ at the level of 0.1 -- 0.3~eV.

\subsection{General description}

The  NEMO~3 detector  is cylindrical  in  design and  is composed of
twenty equal  sectors.  The external  dimensions of the  detector with
shields are about 6~m in  diameter and 4~m in height.  
NEMO~3 is based on the techniques tested on previous incarnations of
the experiment NEMO~1~\cite{DAS91}, and NEMO~2~\cite{ARN95}: 

The  wire chamber is  made of  6180 open  octagonal drift  cells which
operate in Geiger  mode (Geiger cells).  A gas  mixture of $\sim$ 95\%
helium, 4\% ethyl-alcohol, 1\% argon and 0.15\% water at 10~mbar above
atmospheric pressure is  used as the filling gas  of the wire chamber.
Each  drift  cell provides  a  three-dimensional  measurement of  the
charged particle tracks by recording the drift time and the two plasma
propagation  times.  The  transverse position  is determined  from the
drift  time,  while the  longitudinal  position  is  deduced from  the
difference between  the plasma propagation  times at both ends  of the
cathode wires.   The average vertex  position resolution for  the two-electron
events is $\sigma_t = 0.5$~cm  in the transverse plane of the detector
and $\sigma_l = 0.8$~cm in the longitudinal plane.
The Geiger counters information is treated by the track reconstruction
program   based  on  the   cellular  automaton   algorithm,  described
in~\cite{tracking}. 

The calorimeter, which surrounds the wire chamber, is composed of 1940
plastic   scintillator  blocks   coupled  by   light-guides   to  very
low-radioactivity   photomultiplier    tubes   (PMTs)   developed   by
Hammamatsu.   The energy  resolution $FWHM/E$  of  the calorimeter
ranges  from  
14.1 to 17.6\% for  1~MeV  electrons, while  the  time
resolution is of 250~ps at 1~MeV.

The apparatus accommodates almost 10~kg of different double beta decay
isotopes,  comprising  \nuc{100}{Mo}  (6914~g), \nuc{82}{Se}  (932~g),
\nuc{116}{Cd}  (405~g), \nuc{130}{Te}  (454~g),  \nuc{150}{Nd} (34~g),
\nuc{96}{Zr} (9~g),  and \nuc{48}{Ca}  (7~g).  
Most of  these  isotopes are highly enriched,  between 95 --  99\%, and are
produced  in the  form  of thin  metallic  or composite  foils with  a
surface density of 30 -- 60~mg/cm$^2$. 
Three sectors  are also
used for external background  measurement and are equipped respectively
with pure Cu (one sector,  621~g) and natural Te (1.7 sectors, 614~g
of $^{nat}$TeO$_2$).
Some of the sources, including $^{100}$Mo,
have been purified in
order  to  reduce their  content  of  \nuc{208}{Tl} (from $^{232}$Th
and $^{228}$Th, and from $^{228}$Ra with half-life of 5.75~y) and
\nuc{214}{Bi} (from $^{226}$Ra with half-life of 1600~y)
either  by  a  chemical  procedure~\cite{ARN01},  or  by  a  physical
procedure~\cite{ARN05}.  The foils  are placed inside the wire chamber
in the  central vertical  plane of each  sector.  The majority  of the
detector, 12 sectors, is mounted with 6.9~kg of \nuc{100}{Mo}.

The detector is surrounded  by a solenoidal coil which generates
a vertical magnetic  field of 25~Gauss inside the  wire chamber.  This
magnetic field  allows electron-positron separation by measuring
the curvature of tracks.  The ambiguity of the e$^+$/e$^-$ recognition
based on the curvature reconstruction is 3\% at 1~MeV.
 
The  whole detector  is  covered  by two  types  of shielding  against
external  $\gamma$-rays and  neutrons.  The  inner shield  is  made of
20~cm thick  low radioactivity iron  and stops $\gamma$-rays  and slow
neutrons.  The  outer shield, comprised  of tanks filled  with borated
water  on the  vertical walls  and  wood on  the top  and bottom,  was
designed in order to 
thermalise and capture neutrons.

The experiment is located  in the Fr\'ejus Underground Laboratory, the
LSM\footnote{Laboratoire     Souterrain     de    Modane;
\texttt{http://www-lsm.in2p3.fr}},  at  the   depth  of  4800~m  water
equivalent where the  remaining cosmic ray flux is  only 4.2 muons per
m$^2$ per day~\cite{BER87}.  In May  2002 the detector  took its
first  data  which   were  used  for  the  study   of  the  detector's
performance.   Since February  2003,  after the  final  tuning of  the
experimental set-up, NEMO~3 has  routinely been taking data devoted to
double beta  decay studies.  The calibration  with radioactive sources
is carried  out every  6 weeks.  The  stability of the  calorimeter is
checked daily with a laser based calibration system~\cite{ARN05}.

The advantage of the NEMO~3 detector rests in its capability to
identify the two electrons from $\beta\beta$ decay and the
de-excitation photons from the  excited state of the daughter nucleus.
The  NEMO~3  calorimeter  also  measures  the detection  time  of  the
particles.   The  use of  appropriate  time-of-flight  (TOF) cuts,  in
addition  to  energy  cuts,  allows  an  efficient  reduction  of  all
backgrounds.

\subsection{Background}
According  to its  origin,  the  background events  in  NEMO~3 can  be
divided  into  two categories.   The  first  category  is called  {\em
internal  background\/} and  has  its origin  inside  the source  {\em
foils\/}.  This background is  mainly due to the  presence of
radioactive  impurities  in  the  foils.  Particularly  dangerous  are
$\beta^-$-emitters  with  high  $Q_\beta$  values  like  \nuc{208}{Tl}
($Q_\beta  =  4.99$~MeV)  and  \nuc{214}{Bi} ($Q_\beta  =  3.27$~MeV).
These isotopes  can produce $\beta\beta$-like events  through three
mechanisms: (i) a $\beta$-decay  accompanied by an electron conversion
process, (ii) the M\"oller scattering of a $\beta$-decay electron, and
(iii) a  $\beta$-decay emission  to an excited  state followed  by the
Compton scattering of the de-excitation photon.

Great care was taken in  the production and subsequent purification of
the enriched materials, as well  as during the source foil fabrication
itself and its installation into the detector to keep any background 
contamination
to  the  minimum.   The  source  foils were required  to  satisfy  strict
radioactivity    limits~\cite{ARN05}.  

In  addition to  the  radioactive impurities,  the $2\nu\beta\beta  \,
({\rm  g.s.}  \rightarrow {\rm  g.s.})$  decay  is  also a  source  of
internal background  for the $2\nu\beta\beta$ decay  to excited states
signal  because of  brems\-strah\-lung radiation  from  the electrons.
The tail of the $E_{\rm ee}$ distribution from $2\nu\beta\beta$ 
overlaps  the window of 
interest, to some extent, for the $0\nu\beta\beta$  decay  search.   
Its  contribution  depends  on  the
$2\nu\beta\beta$  decay half-life  and  the energy  resolution of  the
detector.

The second category  of background in the NEMO~3  experiment is called
{\em  external background\/}  and is  caused by  electrons  or photons
generated everywhere  else except within the foils.   One component of
this background is due to  radioactive isotopes present 
in  the   detector  construction   materials.   This
contribution was minimised by  a strict selection of low radioactivity
materials during the  NEMO~3 construction \cite{ARN05,BUS02}.  Another
component comes  from the laboratory  walls and from radon  present in
the laboratory air.  Background  from cosmic rays is negligible thanks
to the  placement of  the detector in  the underground  site.  Neutron
background is also negligible due  to the two types of shields against
neutrons and  $\gamma$-rays (from neutron capture) and  because of use
of  the  magnetic  field,  which  during  data  processing  allows  the
elimination of  e$^+$e$^-$ pairs  coming  from the interaction  of 
high  energy photons with the source foils.

The most bothersome external background contribution comes from radon.
Radon is a highly diffusive  radioactive gas and is outgased into the
air  from the  rock walls  of the  LSM laboratory.   It can  enter the
detector either through tiny gaps  between the sectors, or through gas
piping  joints.   The  average  radon  activity  in  air  of  the  LSM
laboratory is  around 10 --  20~Bq/m$^3$, while the measured  value of
average volume activity  of radon inside the NEMO~3  input gas from
February 2003 to  September 2004 was about 25  -- 30~mBq/m$^3$.  
Radon disintegrates to $^{214}$Pb via two $\alpha$-decays. The latter, after a
subsequent $\beta^-$-decay, generates $^{214}$Bi. 
The  decay of \nuc{214}{Bi} to $^{214}$Po is generally
%
%
accompanied by one electron and several photons,  it has a recognisable
signature consistent with a real $\beta\beta$ event
and can only  be partially rejected by the TOF cuts.
$^{214}$Po  is an $\alpha$-emitter (T$_{1/2}$ =
163~$\mu$s) and decays to $^{210}$Pb. NEMO~3 records events with
a delay time up to 700~$\mu$s in order  to tag the
$\alpha$-particles from the $^{214}$Po. 


\section{Analysis and results}

In this work we use the data taken  from March 17, 2003 up to
September 23, 2004. This  time corresponds  to the  stable period  of the
operation  of the NEMO~3 detector  before  the  installation of an
anti-radon factory in order to reduce radon background. The  efficient
data collection
time is 8024~hours (0.92~y) and 
the total number of triggers was 2.2$\cdot$10$^8$.

%
\subsection{Particle identification in NEMO~3}

The NEMO~3  detector is capable of identifying  electrons, positrons,
photons, and $\alpha$-particles by unambiguous topological signatures.
We define them  in the following way:
\begin{itemize}
\item An  electron coming from the source foil is recognised as a
  track going anticlockwise ("negative" track curvature),  found  by
  the track  reconstruction  program  and pointing to a fired
  scintillator. The  scintillator has to be isolated from all other
  fired scintillators. 
\item A positron is defined like an electron but with the opposite
  ("positive") track curvature.
\item  A photon  is  defined as a group of one or more fired
scintillators, in the same region of the detector, 
unassociated with any  track. It is required that all the
scintillators   in  the   group  have to be fired close together
in  time
in order to exclude a random coincidence of hits of different origin.
\item An $\alpha$-particle is defined as a group of delayed Geiger
  cell hits,
  which are in time within 1.5~$\mu$s and close to each
  other in space.  It is required that an  $\alpha$-particle has to
  be close to 
  the event vertex or to the electron (positron) track.
\end{itemize}

\begin{figure}[tb]
\centerline{
\parbox[t]{7.0cm}{\epsfxsize7.0cm\epsffile{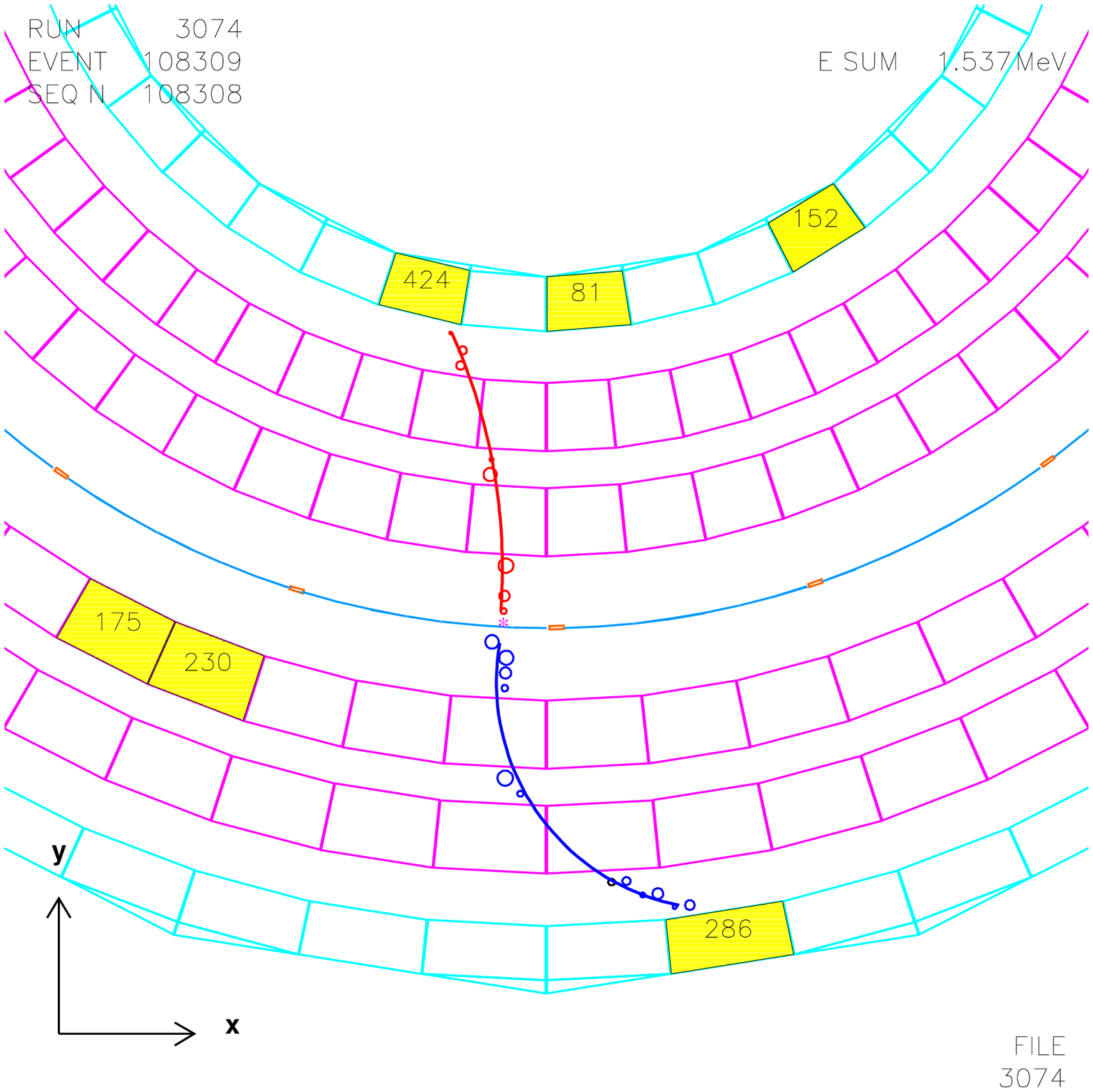}}
\parbox[t]{7.0cm}{\epsfxsize7.0cm\epsffile{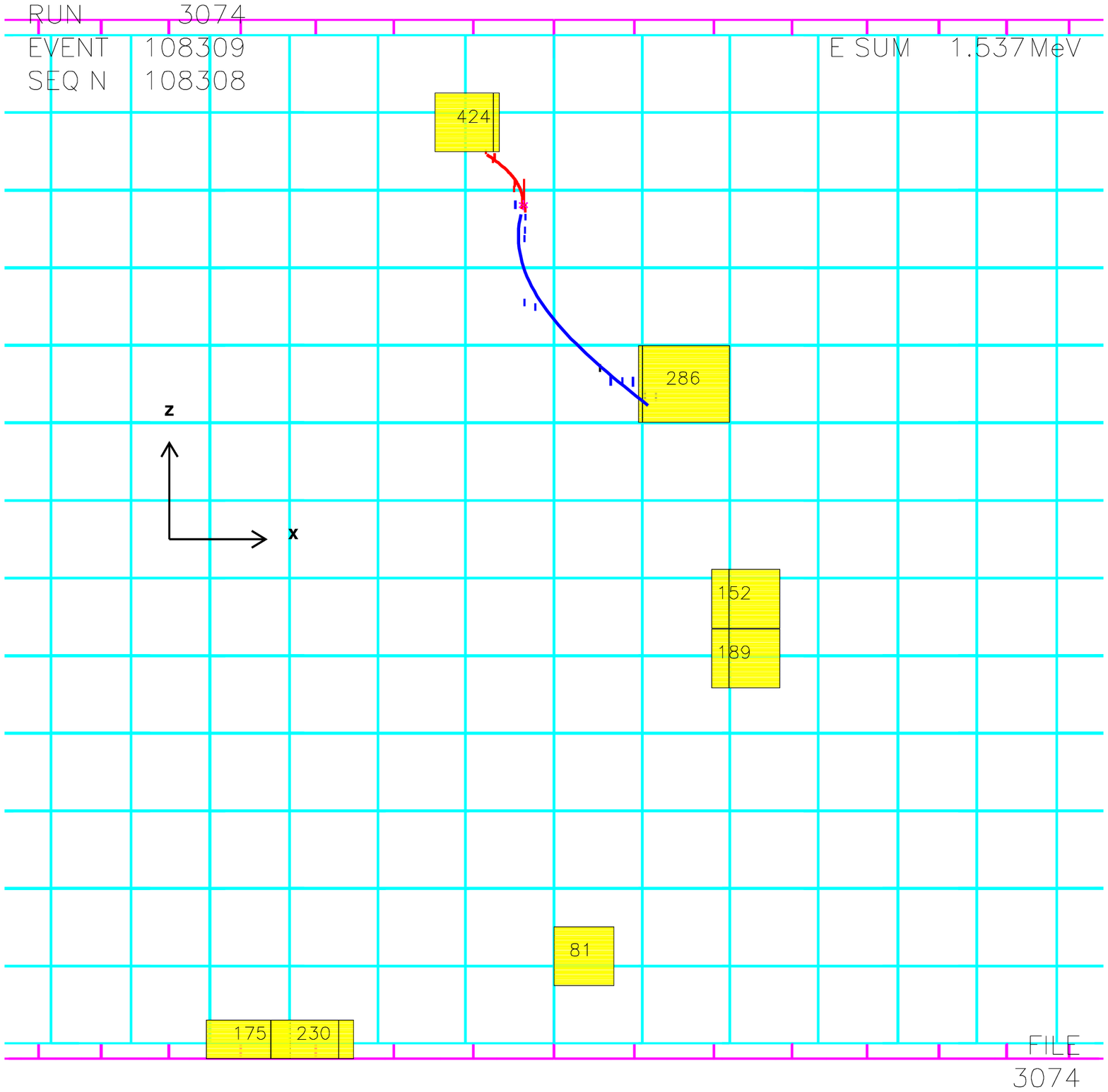}}}
\setlength{\unitlength}{1mm}
\vspace*{-7.6cm}
\begin{picture}(145,75)
\linethickness{0.15mm}
\put(1,0){\line(1,0){70}}
\put(1,70){\line(1,0){70}}
\put(1,0){\line(0,1){70}}
\put(71,0){\line(0,1){70}}
\put(72,0){\line(1,0){71}}
\put(72,70){\line(1,0){71}}
\put(72,0){\line(0,1){70}}
\put(143,0){\line(0,1){70}}
\put(20,70){\makebox(35,5){TOP VIEW}}
\put(90,70){\makebox(25,5){SIDE VIEW}}
\end{picture}
\caption{Top  (left)   and  side  (right)  views   of  a  reconstructed
event candidate selected from data. Two electrons of 424
and  286~keV energy hit  fired isolated scintillators.   They have  a common
origin in  the source foil.  The remaining  fired scintillators compose
three $\gamma$-clusters of  405 (175+230)~keV, 341  (152+189)~keV and
81~keV.   The two clusters of highest  energy are due to
$\gamma$-rays  emitted in coincidence  with the electrons.   The third
cluster  is most  probably due  to the  rescattering of  one  of these
$\gamma$-rays.  In this event  the total deposited energy is 1537~keV.
}
\label{fig:event_example}
\end{figure}

An example of a typical real event, selected for the study of the double
beta decay of $^{100}$Mo to the  excited 0$^+_1$ state, is shown in 
Fig.~\ref{fig:event_example}.

\subsection{Background estimation methods}

Two  different methods  of background estimation  are used here to
evaluate  the  decay  half-life:  the first  one  is based  on
Monte-Carlo (MC)
simulations (Method~1)~\cite{ValaThese}, while the second one is based
on event analysis in  
selected non-Mo\-lyb\-de\-num foils (Method~2).


The Method~1 requires  the knowledge of  activities  of all the
background sources.  These activities have  been measured by  the HPGe
detectors in the LSM laboratory
and by  radon detectors~\cite{ARN05}.  A special data
analysis allowed the  verification of  these radioactive 
contamination levels and further,
to determine them with higher precision.


The presence  of non-Molybdenum foils  in the NEMO~3  set-up allows them
to be used for  independent  background estimation  and  to provide 
a cross check (Method~2) to  the result  obtained with the 
Method~1. 
The foils made of Copper, natural and enriched Tellurium, Selenium
and Cadmium  are used to this  end.  It is supposed  that the
contribution from double  beta decay  to the excited states of other
sources such  as Tellurium,  Selenium  and Cadmium 
can be  neglected due to the  greater predicted   half-lives    than
for Molybdenum~\cite{SUH98,Domin05}.  Consequently,  the events selected in
the non-Molybdenum foils  are expected to be composed  only of the sum
of the contributions of the internal foil contaminations specific for each
source and of the common external background,  where the main
background component 
is  due  to radon  in  the  tracking wire chamber.  Knowing  the  internal
contamination  values  for   each  source  foil~\cite{ARN05},  one  can
evaluate    the   contribution    corresponding   to    the   
radon
background.   The  number  of   events  due   to  the   internal  foil
contamination evaluated  by MC calculation is found  to be negligibly
small compared to the number due to the 
radon background.
The number of
events from the radon background in non-Molybdenum foils is used to
evaluate the radon background  in Molybdenum by scaling
in proportion to their foil surfaces.
The
correction due  to the variation of the  event registration efficiency
for  different detector  parts  has been  also  applied.  The
recalculation coefficient is determined to be equal to 1.7.

For Method~2 it is not necessary to  know the absolute level
of the  main background  from MC  simulation. 
This is an advantage  of the method.  However its accuracy is
limited  by low  statistics of  observed events  in non-Molybdenum
foils.  

\subsection{Event selection}
\label{sect:cut1}
In this section we described just the common selection criteria for the
four channels of analysis. Other criteria specific for the different channels
are given in the section devoted to the each analysis (see
Secs.~\ref{sect:2b2n01+}, \ref{sect:exs0_0nu}, \ref{sect:2b2n21+} and
\ref{sect:2b0n21+}). 

Only events containing  two electrons
coming from  the source  foils and each depositing energy  greater than
200~keV in the scintillators are selected.  It is required
that each track be associated with a unique scintillator which must 
have  a gain greater  than 0.17~(keV/ADC channel)$^{-1}$.  
Only  the tracks starting from the  first two Geiger
cell layers are kept.  The track intersection with the  source foil is
called a track  vertex and then the event  vertex, i.e. the presumed
double  beta decay  origin  in  the source  foil,  is the  average
position of the  track vertices. Each event has to  have its vertex on
the source  foil.  
The distance between the track vertex
and the event vertex is set to be  within 2~cm in the drift plane and
within 
4~cm  in the  vertical direction. In  order to prevent  the wrong
determination  of the  charge of shorter tracks,  the events  with a  track
associated with the  scintillators on the top or bottom
endcaps of the detector
close to the source foils are rejected.

To determine whether the origin of the electrons in the event 
is the source foil (``internal'') or outside the foil (``external''), 
the time of flight of both
particles is used to calculate 
the corresponding probability of each hypothesis
$P_{\rm int}$ and $P_{\rm ext}$~\cite{ARN05}.  The event is accepted as a double 
beta decay candidate  if $P_{\rm  int} > 0.01$ and $P_{\rm ext} < 0.001$.
 
According  to the  decay scheme  the number  of  $\gamma$-rays simultaneously
emitted from the  event vertex (``internal  $\gamma$-ray'') must be
exactly two for the decay  to the 0$^+_1$  state and exactly one  for the
decay to the 2$^+_1$ state. The estimation of the probability that the
$\gamma$-ray was  emitted together with  each electron from  the source foil
({\em internal  hypothesis})   or  crossing  the   tracking
chamber  ({\em external hypothesis}) was  based on  the time-of-flight
information.  Only events 
with  probabilities  for electron-$\gamma$  pairs  $P_{\rm int}
> 0.01$  and $P_{\rm ext} < 0.01$ are selected.

In order to  suppress the background contribution due  to the radon in
the  tracking chamber, events with 
$\alpha$-particles must be excluded.
The tagging of $\alpha$-particles allows the reduction of the  radon
background contribution by $\approx$40\%.


\subsection{$2\nu\beta\beta$ decay of  $^{100}$Mo to the 0$^+_1$ excited
state of $^{100}$Ru}
\label{sect:2b2n01+}
\begin{table}[bt]
\caption{The relative fraction of events with different numbers of
  $\gamma$-clusters for  the data and MC simulation 
  of the signal  and  the main background,  radon,  in the tracking
  chamber inside NEMO~3.} 
\label{tab:t1}

{\small
\begin{center}
\begin{tabular}{|r|c|c|c|}\hline
\multicolumn{1}{|p{2.5cm}}{Number of $\gamma$-clusters per event}&
\multicolumn{1}{|p{3cm}}{Experiment}&
\multicolumn{1}{|p{3cm}}{MC simulation of $2\nu\beta\beta$
(0$^+\rightarrow$0$^+_1$) of $^{100}$Mo}&
\multicolumn{1}{|p{3cm}|}{MC simulation, radon background} \\ 
\hline
1 & 24.6\% & 15.6\% & 24.1\% \\
2 & 56.7\% & 56.2\% & 55.2\% \\
3 & 14.4\% & 21.5\% & 16.9\% \\
$>$3 & 4.3\% & 6.7 \% & 4.4\% \\ \hline
\end{tabular}
\end{center}
}
\end{table}

Since in the double beta decay to the 0$^+_1$ excited state there are
two  electrons and two $\gamma$-rays,  it is required that the number of
$\gamma$-clusters must be greater or equal to two.  This selection
criterion is motivated by the 
fact that a $\gamma$-ray emitted from the source foil can be scattered off a
scintillator  depositing  in  it   some  energy  and  then  hit  another
scintillator. This  requirement allows us to increase the  efficiency of
the  signal   selection.  The  corresponding   information  about  the
proportion  of events with  one, two  or more  $\gamma$-clusters  for the
data and MC is given in Table~\ref{tab:t1}.

According to the decay scheme, the energy  of each emitted $\gamma$-ray
can not be greater than 590~keV. In order to obtain the biggest
signal-to-background ratio (S/B),  the upper limit on $\gamma$-cluster
energy is set to 550~keV and, at the same time,  the energy of a
cluster corresponding to the ``internal  $\gamma$-ray'' cluster has to
be  greater  than  100~keV.  However,  if  the energy  of  an  extra  fired
scintillator which is not an ``internal photon''  is   more  than
150~keV, the event   is  rejected.  This 
requirement  allows us  to  suppress  events  originating  from  an
external $\gamma$-ray that  deposits a part of its  energy in one scintillator
and then interacts with the source foil imitating the signal process.

\begin{figure}
\centerline{
\parbox[t]{14.0cm}{\epsfxsize14.0cm\epsffile{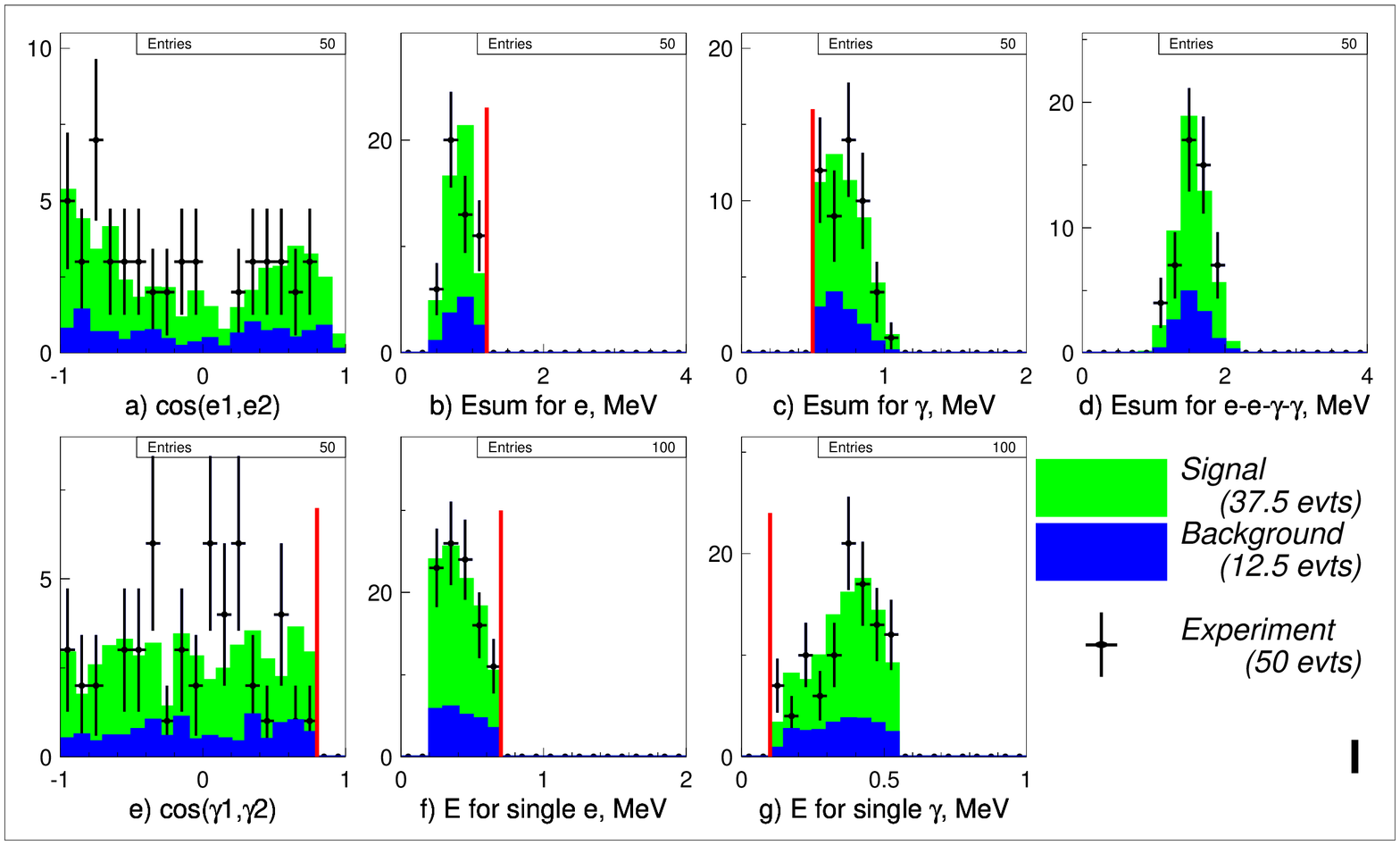}}}
\vspace*{0.5cm}
\centerline{
\parbox[t]{14.0cm}{\epsfxsize14.0cm\epsffile{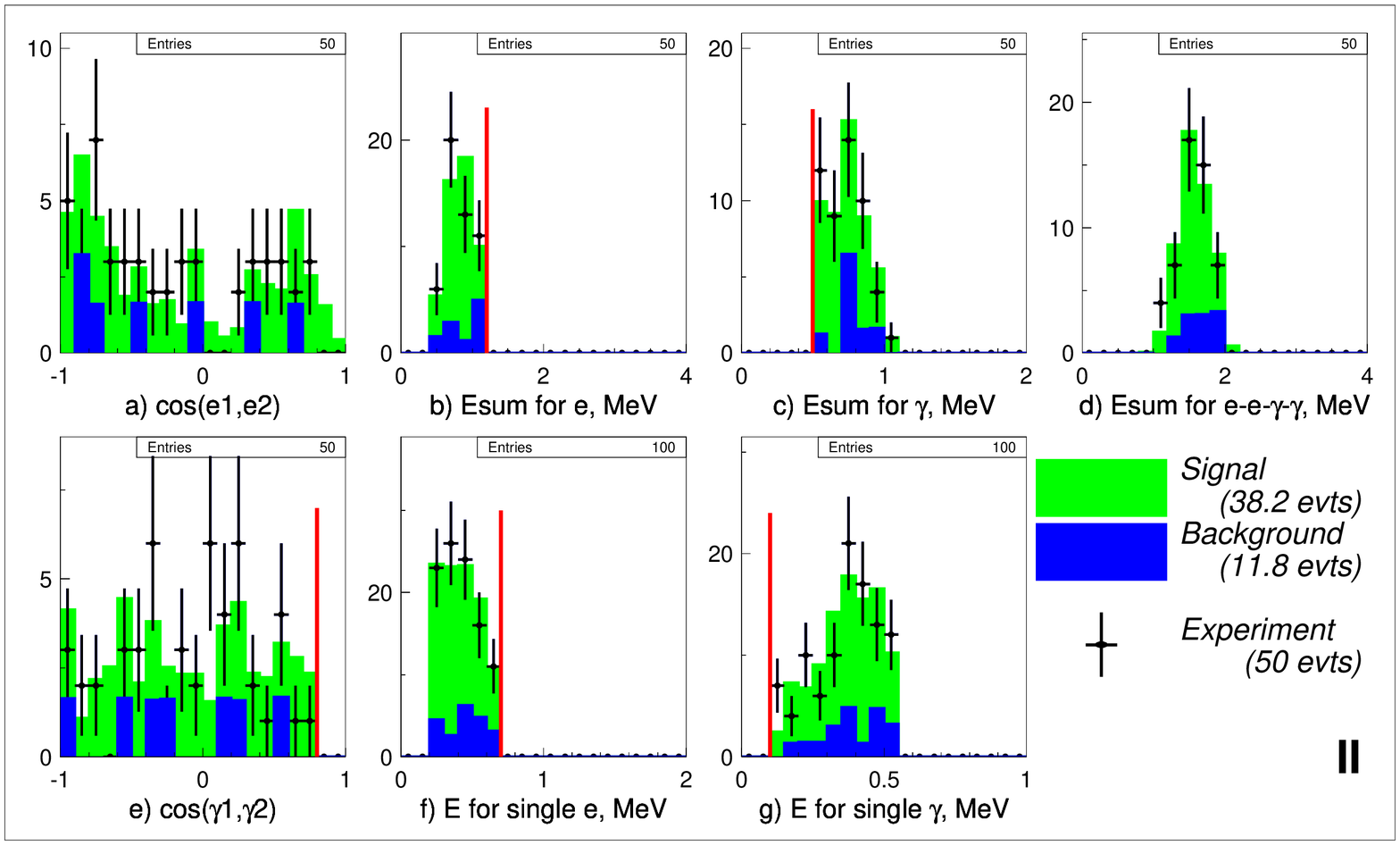}}}
\caption{The main  characteristics of selected events  for the  study of
  2$\nu\beta\beta$ decay of $^{100}$Mo to the 0$^+_1$ excited state of
  $^{100}$Ru after the final
kinematic cut application for Method~1 (I) and for 
Method~2  (II): (a)  the  distribution  of  the cosine of the  angle
between the two electrons,  (b) the energy sum of the two electrons, (c) the
energy  sum  of  the two  $\gamma$-rays,  (d)  the total energy  sum, (e)
the distribution  of the cosine of the  angle between the 
two $\gamma$-rays, (f)  the energy of each electron,  (g) the energy
of each $\gamma$-ray. The experimental  points are shown in black, the
MC simulation of the background in blue, the simulation of
the signal process in green.  The fixed positions of the kinematic cuts
for the best result are presented with red lines (E$^{\rm max}_{\rm
  ee}$ = 1.2~MeV, E$^{\rm max}_{\rm e}$ = 0.7~MeV,   E$^{\rm
  min}_{\gamma\gamma}$  =  0.5~MeV,   E$^{\rm min}_\gamma$  = 0.125~MeV).}
\label{fig:res_mc}
\end{figure}

160  events remained in the data after this event selection. The total
number of background events defined by Method~1 and Method~2 is found to
be 71.4 and 68.4 events, respectively.  
\begin{figure}[tb]
\centerline{
\parbox[t]{15.0cm}{\epsfxsize15.0cm
\epsffile{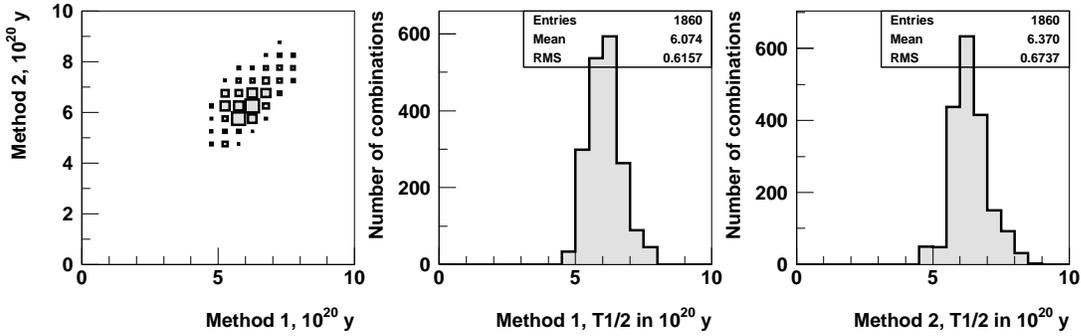}}}
\caption{Half-lives  (T$_{1/2}$) of the double beta  decay of $^{100}$Mo
to the  0$^+_1$ excited state  of $^{100}$Ru, obtained by the two analysis
methods for 1860 combinations of kinematic cuts.}
\label{fig:t12}
\end{figure}
According   to  both
background estimation methods  the background  is mostly due to
the  radon  in the  tracking  volume  of  NEMO~3.  To  suppress  this radon
component in the background,  events with the cosine of the angle between
the two photons  less  than 0.8  are selected
(Fig.~\ref{fig:res_mc}(e)). 
A total of 1860
combinations  of   cuts  on the  following   kinematic 
parameters has been carried out:  
the upper limit  for the energy  sum  of the two  electrons
(E$^{\rm max}_{\rm ee}$): from 1000~keV to  1900~keV in steps of 100~keV; the
upper limit on the energy of  each electron (E$^{\rm max}_{\rm e}$): from
600~keV to 
1000~keV in steps of 50~keV; the  lower limit on the energy  sum for two
``internal   $\gamma$-rays''  (E$^{\rm min}_{\gamma\gamma}$):  from
300~keV  to 800~keV in steps  of  100~keV; the  lower  limit on
the energy of  each  ``internal  $\gamma$-ray'' (E$^{\rm min}_\gamma$):
from  125~keV to  250~keV in 
steps of 25~keV.   For each combination of  cuts the two methods of
background estimation are used.
For each, 
the results of  the calculations of
the half-lives for both analysis  methods as well as the correlation between
them are shown in Fig.~\ref{fig:t12}. The two methods give almost the
same    average    value    of    the   half-life     around
6$\cdot$10$^{20}$~y and are in a very good agreement (see
Fig.~\ref{fig:t12}).

Finally, after the cut optimisation we  have chosen the kinematic
cuts providing the  maximal value of the product
$S/B\times\varepsilon$ using Method~1, where $\varepsilon$ is the
signal registration efficiency.  The optimal set of cuts 
for this criterion     
is given  in
Table~\ref{tab:results}.  

\begin{table}[tb]
\caption{The results for the 
$2\nu\beta\beta$ decay to the 0$^+_1$ excited
  state: the  
signal registration efficiency
  ($\varepsilon$), signal-to-background  ratio (S/B),  total number
  of selected  events (S+B) and statistical  significance of the
  signal (N$\sigma$  = S/$\sqrt{S+B}$) for the two analysis
  methods. Results for both methods are given for the same kinematic
  cuts on the variables: 
E$^{\rm max}_{\rm ee}$, 
E$^{\rm   max}_{\rm e}$,  
E$^{\rm min}_{\gamma\gamma}$, 
E$^{\rm    min}_\gamma$. 
} 
\label{tab:results}
{\small
\begin{center}
\begin{tabular}{|c|c|c|c|c|c|c|c|c|}\hline
method&\multicolumn{4}{|c|}{Kinematic cuts in keV}&
$\varepsilon\cdot$10$^{-4}$&S/B&S+B&N$\sigma$\\ \cline{2-5}
&E$^{\rm max}_{\rm ee}$&E$^{\rm max}_{\rm e}$&E$^{\rm
  min}_{\gamma\gamma}$&E$^{\rm min}_\gamma$&
&&&\\ \hline\hline 
Method~1&1200&700&500&125&
8.1$\pm$1.0&3.0&50&5.3\\
Method~2&1200&700&500&125&
8.1$\pm$1.0&3.2&50&5.4\\[0.1cm]\hline
\end{tabular}
\end{center}
}
\end{table}

The distributions of  the decay characteristics of selected
events for Method~1 after applying optimal kinematic cuts  are
shown in Fig.~\ref{fig:res_mc}(I) 
and the   corresponding  background  component  contributions
evaluated  by this  method are  shown in  Table~\ref{tab:t3}.  In
Fig.~\ref{fig:res_mc}(II) the same spectra obtained using 
Method~2 are  present. 
The numbers of signal and background events after the cuts obtained
for Method~1 are:  37.5$\pm$7.1(stat.)$\pm$5.2(syst.) and
12.5$\pm$2.9(syst.). For Method~2   the numbers are
 38.2 $\pm$ 8.4(stat.) $\pm$ 4.7(syst.) and
 11.8 $\pm$ 4.5(stat.) $\pm$ 0.4(syst.).

The estimation of the systematic uncertainty has been made taking into
account the following contributions: 
\begin{itemize}
\item The  two electron detection inefficiency  due to the
  inaccuracy  of the  GEANT simulation and inaccuracy of the tracking
  program is 
  estimated to be less than 5\%. It is determined by reproducing the
  activity of the calibrated source placed inside the NEMO~3 detector. 
\item The  efficiency calculation error due to the
  inaccuracy  of the GEANT simulation for $\gamma$-rays  is found by
  comparing the MC simulation and data from the
  calibration source. For the  two $\gamma$-channel it is  estimated to be
  around 10\%.  
\item  The  effect  of  the  uncertainty  in  the  energy  calibration
coefficients causes a 2\% contribution  to the systematic  error in
the  half-life.   It  has been  estimated  by  the energy  scale
variation for the experimental data.
\item The effect of the single electron energy threshold for the two
  electron signal is estimated as 3\%. 
\item The effect of the uncertainty of the detector efficiency 
is evaluated from the comparison of results of
the $2\nu\beta\beta$ decay to the ground state separately for
metallic and composite foils of $^{100}$Mo. It is estimated to be around 4\%.
\item  The uncertainty  of the  radon level  measurement is  20\% which
gives a 6\%  error in the half-life. This  uncertainty is inherent
only in analysis Method~1. All other background uncertainties do not
introduce any significant error into the evaluation of result.
\end{itemize}

\begin{table}[tb]
\caption{The calculated  number of events in  the observed spectrum for
$^{100}$Mo foils from each background component. Only the components
leading to a number of expected events greater than 0.1 are given here.}
\label{tab:t3}
{\small
\begin{center}
\begin{tabular}{|p{2.cm}|c|c|c|c|c|c|c|}\hline
Background component &
\multicolumn{1}{|p{1.3cm}|}{$2\nu\beta\beta$ (to g.s.)}&
\multicolumn{1}{|p{1.3cm}|}{internal $^{214}$Bi}& 
\multicolumn{1}{|p{1.3cm}|}{internal $^{208}$Tl}& 
\multicolumn{1}{|p{2.5cm}|}{$^{214}$Bi (radon inside tracking chamber)}&
\multicolumn{1}{|p{2.5cm}|}{$^{214}$Bi (radon outside tracking
  chamber)}\\ \hline   
Number of events& 
0.5& 0.2&  0.3&  11.2&  0.3 \\ \hline 
\end{tabular}
\end{center}
}
\end{table}  
The total systematic error  on the result of analysis Method~1 is
13.8\%,   and    of   the   Method~2   is  
12.4\%.  It  is comparable  with the spread  of the  half-lives
for the  1860  kinematic   cut  combinations  presented   in  
Fig.~\ref{fig:t12}.

The final results  of the half-life  measurements for the two neutrino
double beta decay of $^{100}$Mo to the 0$^+_1$ excited state of
$^{100}$Ru for both methods are the following:
\[
T^{(2\nu)}_{1/2}(0^+\rightarrow0^+_1) =
5.7^{+1.3}_{-0.9}(stat.)\pm0.8(syst.)\cdot10^{20}\; {\rm y \;(Method\;1)};
\] 
\[
T^{(2\nu)}_{1/2}(0^+\rightarrow0^+_1) =
5.6^{+1.5}_{-1.0}(stat.)\pm0.7(syst.)\cdot10^{20}\; {\rm y \;(Method\;2)}.
\] 


\subsection{Search for the 0$\nu\beta\beta$ decay to the 
0$^+_1$ excited state of $^{100}$Ru}
\label{sect:exs0_0nu}
For the study  of the neutrinoless double beta  decay of $^{100}$Mo to
the 0$^+_1$ excited  state of $^{100}$Ru, in addition  to the selection
criteria  described  in  Sec.~\ref{sect:cut1},  
the
energy of  each electron is required to be  less than  1600~keV and the 
energy of each photon ($E_\gamma$) to be greater than 125~keV and less than
600~keV according to  the particle energy  distributions for the
0$\nu$  mode.

\begin{figure}[tb]
\centerline{
\parbox[t]{12.0cm}{\epsfxsize12.0cm
\epsffile{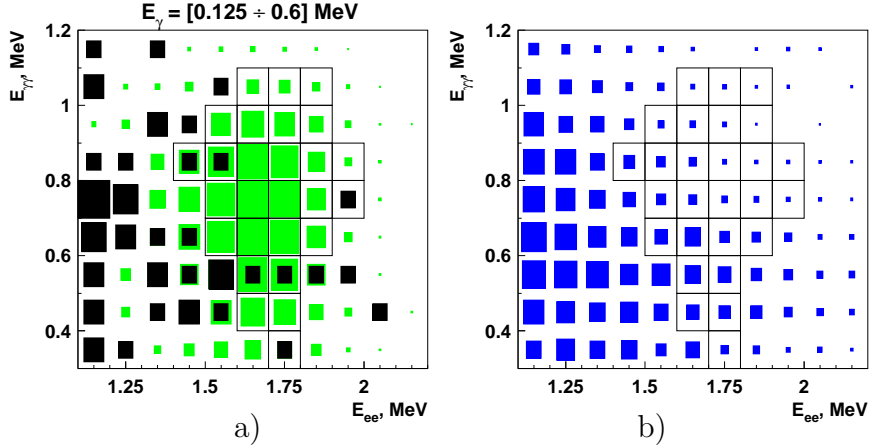}}}

\vspace*{-0.5cm}\hspace*{4cm}a)\hspace*{5cm}b)
\caption{Scatter plot of the sum of the $\gamma$-rays energy versus the
energy sum of the electrons a) for the data (in black) and b) for the MC
background (in blue). The  simulated $0\nu\beta\beta$ to  the 0$^+_1$
excited state signal ((a), in green)  is also shown with an arbitrary  scale.
The bins with $\varepsilon_S/\varepsilon_B$ $>$
$\langle\varepsilon_S/\varepsilon_B\rangle$ are shown with a black frame.}
\label{fig:exs0_0nu}
\end{figure}
The two-dimensional distribution of the
energy   sum  of the two $\gamma$-rays   ($E_{\gamma\gamma}$) and  the energy
sum of the two electrons  ($E_{\rm ee}$) is shown in Fig.~\ref{fig:exs0_0nu}.
The expected signal ($\varepsilon_S$)  and the background
($\varepsilon_B$) efficiencies can be found for each bin of this
distribution from the MC simulation.   
The signal-to-background ratio 
($\varepsilon_S/\varepsilon_B$) may be calculated for each bin,
and their average value $\langle\varepsilon_S/\varepsilon_B\rangle$
among all bins is found. 
It is reasonable to 
consider only the bins with a  significant $\varepsilon_S$.
If 
$\varepsilon_S/\varepsilon_B$ $>$
$\langle\varepsilon_S/\varepsilon_B\rangle$ 
is required, 
the region searched is composed of 27 bins (see the Fig.~\ref{fig:exs0_0nu}). 
6 experimental events are observed and 6.6 MC background events are 
expected.
Comparing experimental data with the MC background there is no
evidence for the 0$\nu$ signal. 

To set the limit we apply a method based on the confidence level
computation with small statistics  (CLs method) 
~\cite{TJunk} treating each 
bin of the distribution as independent search channel. 
The uncertainty in the number of background events
has been set at 20\% (mostly due to the radon level
uncertainty), for the signal efficiency error we use 5\%.
Using the MCLIMIT program~\cite{MCLimit} the number of signal events 
excluded at  90\% confidence level is
calculated to  be 3.9. Taking into account the signal efficiency of 1.3\%
one may evaluate the limit on the 0$\nu$ decay half-life:
\[
T^{(0\nu)}_{1/2}(0^+\rightarrow0^+_1) > 8.9\cdot10^{22} {\rm y \;\;(at\;\; 
90\%\;\;C.L.)}. 
\]
\subsection{Search for the $2\nu\beta\beta$ decay of $^{100}$Mo to the
2$^+_1$ excited state of $^{100}$Ru}
\label{sect:2b2n21+}

This decay produces the emission  of two electrons  and one
photon  of 540~keV (Fig.~\ref{fig:mo100_decay_scheme}).  To study
this decay channel we perform the same data selection as for the decay
to the  0$^+_1$ excited  state but with the modified criterion that the
number of fired scintillators in event is required to be greater than 
or equal to 3.   The  energy sum  of  all  
$\gamma$-clusters must be less  than
550~keV.  In order to select  events in the energy region specific for
this decay  channel, we  used the following  kinematic cuts  for event
selection: the total  detected energy must be greater  than 1000~keV; the
energy of  the ``internal  $\gamma$-ray'' must be  greater than
250~keV; the cosine angle between  an electron and the $\gamma$-ray must
be  less than 0.8. The latter  cut is used in order to suppress the possible
background from the double beta decay to the ground state
accompanied by a $\gamma$-ray tangent to one  of electrons emitted
by the Bremsstrahlung process  in the source foil.

531 events result from this selection, 465 of them can be
attributed to
the background  according to analysis Method~1. 
The   main  background
contributions according to  the MC simulations come from  the radon in
the tracking chamber (38\% of  total background), 2$\nu\beta\beta$ decay to
the  0$^+_1$ excited  state (15\%),  2$\nu\beta\beta$  decay to  the ground
state  (15\%),   radon  outside   the tracking chamber (12\%)
and  internal radioactive contamination of PMTs in $^{232}$Th (12\%) and
$^{226}$Ra (12\%). 
The  error on the signal ($\sigma$) is 43.2 and includes
also the systematic  uncertainty of the background  calculation. The
statistical significance  of the disagreement between MC and data is
1.5$\sigma$.
Using the  signal efficiency defined by
the MC simulation (0.45\%)
the  half-life limit of  the 2$\nu\beta\beta$ decay  of
$^{100}$Mo  to the 2$^+_1$ excited state of $^{100}$Ru is calculated to be:
\[
T^{(2\nu)}_{1/2}(0^+\rightarrow2^+_1) > 1.1\cdot10^{21}\; {\rm y
  \;\;(at\;\;  90\% \;\;C.L.)}.
\]

\subsection{Search for the 0$\nu\beta\beta$ decay of $^{100}$Mo to the
2$^+_1$ excited state of $^{100}$Ru}
\label{sect:2b0n21+}
A similar analysis to that described  in Sec.~\ref{sect:exs0_0nu} has
been performed to search for the neutrinoless double beta decay of
$^{100}$Mo to the 2$^+_1$ excited state.  After the preliminary
analysis the events with the typical topology 
were selected and
the two-dimensional distribution 
of the energy sum of the electrons $E_{\rm ee}$ and the energy of
the $\gamma$-rays $E_{\gamma}$ was analysed.

\begin{figure}[tb]
\centerline{
\parbox[t]{12.0cm}{\epsfxsize12.0cm
\epsffile{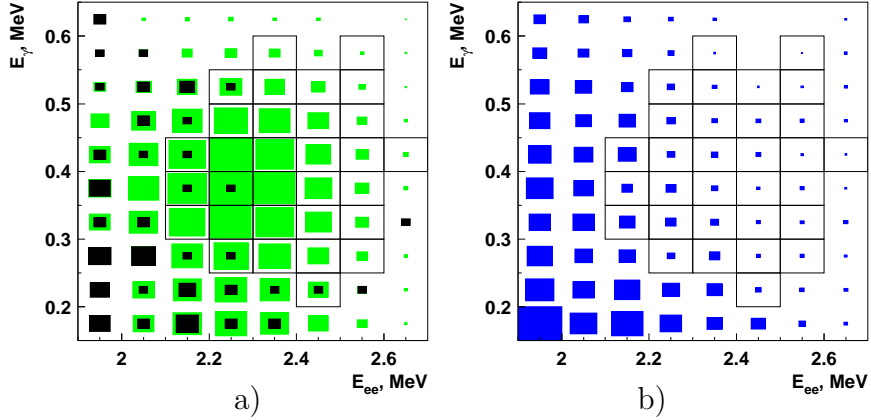}}}

\vspace*{-0.5cm}\hspace*{4cm}a)\hspace*{5cm}b)
\caption{Scatter  plot of the $\gamma$-ray energy  versus the energy sum
of electrons a) for the  experimental data (in black) and b) for MC background
(in blue).  The simulated
0$\nu\beta\beta$ to the 2$^+_1$  excited state signal in arbitrary scale
((a), in green) is  also shown.   The relevant bins are shown with a black frame.}
\label{fig:exs2_0nu}
\end{figure}

Using the same approach as for the 2$\beta0\nu$ (0$^+\rightarrow0^+_1$)
(see sect.~\ref{sect:exs0_0nu}) there are 31 
bins of the distribution with $\varepsilon_S/\varepsilon_B$ ratio
greater than the average as shown in Fig.~\ref{fig:exs2_0nu}.  
The signal efficiency is found to be 2.9\%
\footnote{The N*  mechanism of  $\beta\beta$ decay 
\cite{Doi1981} has  been used for MC simulation.}, 6
events are observed and  
5.7 background  events are  expected.  In
the limit calculations the background   uncertainty  (20\% mostly due
to the radon level uncertainty) and the error on the signal efficiency
(5\%) have   been   taken  into account. 
4.9 signal events are excluded on 90\%~C.L. and the corresponding
limit on the half-life is: 
\[
T^{(0\nu)}_{1/2}(0^+\rightarrow2^+_1) > 1.6\cdot10^{23}\; {\rm y\;\;(at\;\; 
90\%\;\;C.L.)}.
\]
\section{Discussion}

The results  of this analysis of the 
$\beta\beta$-decay of  $^{100}$Mo to the excited states  of $^{100}$Ru are
presented in Table~\ref{tab:discussion} along with previous 
experimental results  and  theoretical  estimates  for
the half-lives.

For the transition to the 0$^+_1$ excited state of $^{100}$Ru the obtained
value is in a good agreement with results of previous experiments.  It
has to be stressed here  that in the present experiment with the
NEMO~3 detector, all the decay products (two  electrons
and two $\gamma$-rays)  were  detected for the first time and hence 
all the  useful
information  about the decay  was obtained (total energy spectrum, single
electron spectrum, single $\gamma$ spectrum, and all angular  distributions).
In  previous  experiments
either  individual $\gamma$-rays~\cite{BAR95,BAR99}, or the coincidence
between the two emitted $\gamma$-rays~\cite{BRA01}  were  detected  but
no information about electrons was obtained.

\begin{table}
\caption{Experimental   results   and   theoretical  predictions   for the
half-life  of the $\beta\beta$-decay of  $^{100}$Mo to  the excited  states of
$^{100}$Ru. All limits (except one specifically mentioned) are at 90\%
C.L.  Theoretical  predictions  for  0$\nu$  mode  are  presented  for
$\langle   m_\nu\rangle$  =   1  eV   and   $\langle\lambda\rangle$  =
10$^{-6}$.}
\label{tab:discussion}
\begin{center}
\begin{tabular}{|l|l|l|l|}
\hline
Transi-&T$_{1/2}$, y&T$_{1/2}$, y&T$_{1/2}$, y\\
tion&(this work)&(previous works)&(theor. predictions)\\ \hline
$0\nu\beta\beta$;&$>$1.6$\cdot$10$^{23}$&$>$1.4$\cdot$10$^{22}$(68\% C.L.)~\cite{Ejiri96}&6.8$\cdot$10$^{30}$($\langle m_\nu\rangle$)\\
0$^+\rightarrow$2$^+_1$&&&
2.1$\cdot$10$^{27}$($\langle\lambda\rangle$)~\cite{Tomoda2000}\\ \hline
$2\nu\beta\beta$;&$>$1.1$\cdot$10$^{21}$&
$>$1.6$\cdot$10$^{21}$~\cite{BAR95}&3.4$\cdot$10$^{22}$~\cite{Stoica1996}\\
0$^+\rightarrow$2$^+_1$&&&(5.5-24)$\cdot$10$^{25}$~\cite{Hirsh95}\\
&&&10$^{22}$~\cite{Schwieger98}\\
&&&2.1$\cdot$10$^{21}$~\cite{CIV99}\\\hline
$0\nu\beta\beta$;&$>$8.9$\cdot$10$^{22}$&$>$6$\cdot$10$^{20}$~\cite{DAS95}&(7.6-14.6)$\cdot$10$^{24}$\\
0$^+\rightarrow$0$^+_1$&&&($\langle m_\nu\rangle$)~\cite{Simk2001}\\ 
&&&2.6$\cdot$10$^{26}$($\langle 
m_\nu\rangle$)~\cite{Suhonen2002}\\ \hline 
$2\nu\beta\beta$;&[5.7$^{+1.3}_{-0.9}$(stat)$\pm$0.8(syst)]$\cdot$10$^{20}$&
6.1$^{+1.8}_{-1.1}\cdot$10$^{20}$~\cite{BAR95}&(0.7 - 1)$\cdot$10$^{21}$~\cite{Grif92}\\
0$^+\rightarrow$0$^+_1$&&[9.3$^{+2.8}_{-1.7}$(stat)$\pm$1.4(syst)]$\cdot$10$^{20}$~\cite{BAR99} 
&2.1$\cdot$10$^{21}$~\cite{Stoica1996}\\
&&[5.9$^{+1.7}_{-1.1}$(stat)$\pm$0.6(syst)]$\cdot$10$^{20}$~\cite{BRA01}
&1.5$\cdot$10$^{21}$~\cite{Hirsh95}\\
&&$>$1.2$\cdot$10$^{21}$~\cite{NEMO92}&1.5$\cdot$10$^{20}$~\cite{CIV99}\\ \hline
\end{tabular}
\end{center}
\end{table}

One can compare the measured value with the theoretical predictions, which
were obtained  using QRPA-based models~\cite{Stoica1996,CIV99,Grif92},
and the pseudo-SU(3) model~\cite{Hirsh95}.  The
present accuracy of the theoretical prediction is directly connected to
the accuracy of the nuclear matrix elements (NME) calculations and usually
is  a factor of $\sim$4  for the half-life  (factor of  $\sim$2 for the NME). 
In view of this, it can be seen that
there is  a reasonable agreement between the experimental and theoretical
values of the half-life.

Recently    an    estimation    of    the    half-life    value    for the
2$\nu\beta\beta$ (0$^+ \rightarrow$ 0$^+_1$)  transition of  $^{100}$Mo was
carried out using   the 
so-called    single    state   dominance    hy\-po\-the\-sis
(SSDH)~\cite{CIV99,Abad}  and  the  recent  result for  electron
capture (EC)  transition
$^{100}$Tc           $\rightarrow$           $^{100}$Mo~\cite{Garcia}:
T$^{(2\nu)}_{1/2}$(0$^+\rightarrow              0^+_1$)              =
4.2$\cdot$10$^{20}$~y~\cite{Domin05}  (with an  accuracy  of $\pm$~50\%
because  of such  accuracy of  the EC  transition~\cite{Garcia}). This
value  is in  agreement  with our  result  and if  in  the future  the
accuracy  of the  EC transition  can  be improved,  then a  comparison
between experimental and estimated values will provide the possibility
to check more precisely the SSDH prediction.

Using the  phase-space value  G = 1.64$\cdot$10$^{-19}$~y$^{-1}$ (for
g$_A$ =  1.254) and the obtained  half-life value, one obtains  the nuclear
matrix   element  value  (NME)   for  the 2$\nu\beta\beta$ (0$^+\rightarrow
0^+_1$) transition, M(0$^+_1$) = 0.103$\pm$0.011. One can compare this
value  with the NME  value for  the 2$\nu$-transition to  the ground  state of
$^{100}$Ru, which was obtained also with the NEMO~3 detector, M(g.s.)
=    0.126$\pm$0.005    (here    T$^{(2\nu)}_{1/2}$    =
[7.11$\pm$0.02(stat)$\pm$  0.6(syst)]$\cdot$10$^{18}$~\cite{Arn0n} and
G  = 8.9$\cdot$10$^{-18}$~y$^{-1}$~\cite{SUH98}).  One  can see  that
M(g.s.)   is   $\sim$20\%  greater  than M(0$^+_1$).   Nevertheless,
these  values are  very close  and,  taking into  account errors  and
possible uncertainties,  their equality can not be excluded. From
the viewpoint of the theory  it is important to confirm  this difference or to
rule it out,  in particular to ascertain the validity of the SSDH.  
Future measurements
with  the NEMO~3 detector  where results  for both  transitions  (to
the 0$^+$ ground and the 0$^+_1$ excited states) will be improved, will
address this further.

The limit obtained  for the 2$\nu\beta\beta$(0$^+\rightarrow 2^+_1$)  decay is
comparable with the limit from Ref.~\cite{BAR95} but is 1.5 times lower. In
this channel a small excess of events is observed, which probably
can be  explained by  a statistical fluctuation.  Future measurements
are  needed  to  clarify  the  status of  these  extra  events. The most
optimistic theoretical  prediction for $^{100}$Mo 
gives     T$^{(2\nu)}_{1/2}$(0$^+\rightarrow   2^+_1$)   $\approx$
2$\cdot$10$^{21} - 10^{22}$~y~\cite{Schwieger98,CIV99} for this type of decay
which  is not
very  far  from the obtained  limit. However,  the estimation  in the
framework  of the SSDH,  which looks  more reasonable,  gives a higher
value, T$_{1/2}$ = 1.7$\cdot$10$^{23}$~y~\cite{Domin05}.

Concerning  the  bound on the 0$\nu\beta\beta$  decay  of $^{100}$Mo  to
2$^+_1$ and 0$^+_1$  excited states of $^{100}$Ru the limits obtained in the
present work  are better than results of previous experiments by a
factor of $\sim$10  and  $\sim$100,  respectively.  Using  the results  of
theoretical  calculations  from Refs.~\cite{Simk2001} and~\cite{Suhonen2002}
for the 0$\nu\beta\beta$ decay  of  $^{100}$Mo to  the  0$^+_1$ state  of
$^{100}$Ru, one obtains a limit on the effective value of the Majorana mass,
$\langle m_\nu\rangle$  $<$ 9.3 $-$ 12.8~eV and  60~eV, respectively.  It
means that to reach a sensitivity to $\langle m_\nu\rangle$ at the level of
$\sim$ of 0.1~eV, the sensitivity  to the half-life  has to  be  at  the  level
of $\sim$10$^{27}$~y.

The 0$\nu\beta\beta$(0$^+\rightarrow 2^+_1$)  decay had long been 
accepted to be possible only because of the contribution of
right-handed currents and it is not sensitive to the neutrino
mass contribution.  
However, in Ref.~\cite{Tomoda2000} it was demonstrated that the relative
sensitivities of  (0$^+\rightarrow 2^+_1$) decays to the neutrino mass
$\langle    m_\nu\rangle$  and the right-handed currents 
$\langle\eta\rangle$ are comparable to those of 0$\nu\beta\beta$
decay to the ground state of the daughter nuclei. 
At the same time, the
(0$^+\rightarrow 2^+_1$) decay is relatively more sensitive to
$\langle\lambda\rangle$. Using the
limit   of T$_{1/2}$    $>$   1.6$\cdot$10$^{23}$~y,  the  value for the NME
from Ref.~\cite{Tomoda2000}  and  the phase-space value  from
Ref.~\cite{Tomoda1991}, 
one    can   obtain    the   limits    $\langle    m_\nu\rangle$   $<$
6.5$\cdot$10$^3$~eV       and       $\langle\lambda\rangle$       $<$
1.2$\cdot$10$^{-4}$.

\section{Conclusion}

The results from the NEMO~3 detector for $\sim$1 year
of  measurement and with a higher level of  background from
$^{222}$Rn than expected, have been obtained.  
Data taking will be continued for a further $\sim$5 years with an improved 
level of radon background.   It means that the sensitivity of
the experiment will be  improved  by a factor of  $\sim5-10$.  For
example,  in  the  case  of 
2$\nu\beta\beta$(0$^+\rightarrow 0^+_1$) decay  $\sim$ 200 events will
be  detected  under  conditions  of a signal-to-background  ratio
of $\sim$  10 and  an accuracy for the half-life value
reduced  by $\sim$10\%. The sensitivity  to the
2$\nu\beta\beta$(0$^+\rightarrow  2^+_1$) decay will  be on the level
of $\sim$  10$^{22}$~y and it could also yield a
chance to detect this  type of decay (see Table~\ref{tab:discussion}).
For  the 0$\nu$   transitions  to  the 2$^+_1$  and   0$^+_1$  excited  states
the sensitivity will be $\sim$ 10$^{24}$~y.



\begin{ack}
The authors would like to thank 
the Fr\'ejus Underground Laboratory staff
for their technical assistance in running the experiment.
A portion of this work has been supported by 
the Grant Agency of the Czech Republic 
(grant No.~202/05/P293). 
Another portion of this work has been supported by 
a grant from INTAS (No.~03051-3431), 
by a NATO grant (PST.CLG.980022) 
and by a RFBR (No.~06-02-16672a). 
This work has been partially supported by the ILIAS integrating activity
(Contract No. RII3-CT-2004-50622) as part of the EU FP6 programme
in Astroparticle Physics.

\end{ack}


\end{document}